\documentclass[a4paper,fleqn,usenatbib,useAMS]{mnras}

\usepackage{graphicx}	
\usepackage{amsmath}	
\usepackage{amssymb}	
\usepackage{multicol}        
\usepackage{bm}		
\usepackage{pdflscape}	
\usepackage{epstopdf}	

\def\HI{H~{\sc i}\, }
\def\HII{H~{\sc ii}\, }

\usepackage[T1]{fontenc}
\usepackage{ae,aecompl}

\usepackage{txfonts}

\title[Interacting galaxy NGC4656 and its dwarf companion]{ Interacting galaxy NGC4656 and its unusual dwarf companion \thanks{ Based on observations obtained with the 6-m telescope of the
Special Astrophysical Observatory of the Russian Academy of
Sciences (SAO RAS).}}
\author[A. Zasov et al.]{Anatoly V. Zasov$^{1,2}$\thanks{E-mail:
zasov@sai.msu.ru}, Anna S. Saburova$^1$, Oleg V. Egorov$^1$, Roman I. Uklein$^3$ 
\\
$^1$ Sternberg Astronomical Institute, M.V. Lomonosov Moscow State University, Universitetskij pr., 13,  Moscow, 119234, Russia\\
$^2$ Faculty of Physics, M.V. Lomonosov Moscow State University, Leninskie gory 1,  Moscow, 119991, Russia \\
$^3$ Special Astrophysical Observatory, Russian Academy of Sciences, Nizhniy Arkhyz, Karachai-Cherkessian Republic 357147, Russia \\
}
\begin{document}
\label{firstpage}
\pagerange{\pageref{firstpage}--\pageref{lastpage}} \pubyear{2016}
\maketitle

\begin{abstract}
We studied the nearby edge-on galaxy NGC4656 and its dwarf low surface brightness companion with the enhanced UV brightness,  NGC4656UV, belonging to the interacting system NGC4631/56. Regular photometric structure and relatively big size of NGC4656UV allows to consider this dwarf galaxy as a separate group member rather than a tidal dwarf. Spectral long-slit observations were used  to obtain  the  kinematical parameters and gas-phase metallicity of NGC4656UV and NGC4656. Our rough estimate of the total dynamical mass of NGC4656UV allowed us to conclude that this galaxy is the dark-matter dominated LSB dwarf or ultra diffuse galaxy. Young stellar population of NGC4656UV, as well as strong local non-circular gas motions in NGC4656 and the low oxygen
gas abundance in the region of this galaxy adjacent to its dwarf companion, give evidence in favour of the
 accretion of metal-poor gas onto the discs of both galaxies.

\end{abstract}


\section{Introduction}\label{intro}

Tidal interaction or merging of gas-rich galaxies often lead to the nascence of local sites of star formation in tidal structures in  local regions of enhanced density of gas lost by galaxies.  Under certain conditions gravitationally bound tidal dwarf galaxies (TDGs) may be  formed in the tidal debris (most often -- in the tidal tails). Their distinguishing features are the young stellar population, a  low, if any, content of dark matter,  and the moderate gas metallicity, because they consist of gas expelled from the peripheral regions of more massive parent galaxies.  Some  non-bound and short lived sites of star formation may also appear beyond the main bodies of galaxies as a result of interaction. In addition, a close encounter or collision of galaxies may inspire a gas outflow and/or gas exchange which in turn  may influence the properties of interstellar medium and  gas abundance not only of the interacting galaxies, but also of their dwarf satellites.  

%

The current paper is devoted to the study of the late type galaxy NGC4656, interacting wth NGC4631, and its unusual dwarf companion -- NGC4656UV.
A system NGC4631/56 consists of two moderate-size interacting galaxies \citep{Combes1978}  separated by about a half degree (approximately 65 kpc),  containing the regions of intense star formation, evidently triggered by interaction. Both galaxies are late-type and gas-rich systems observed nearly edge-on. Observations of \HI reveal a complex character of gas dynamics and gas distribution in galaxies with the far outflowing extraplanar streams of low density \HI from NGC4631 \citep{Rand1994}.  The optical tidal stream and several low surface brightness dwarf galaxies were also discovered in the vicinity of NGC4631 \citep[][]{Karachentsevetal2014,  Martinez-Delgadoetal2015}. 

\begin{figure*}                                                              
 \centering    
\vspace{-2cm}  
\hspace{-2.5cm} 
\includegraphics[scale=0.32]{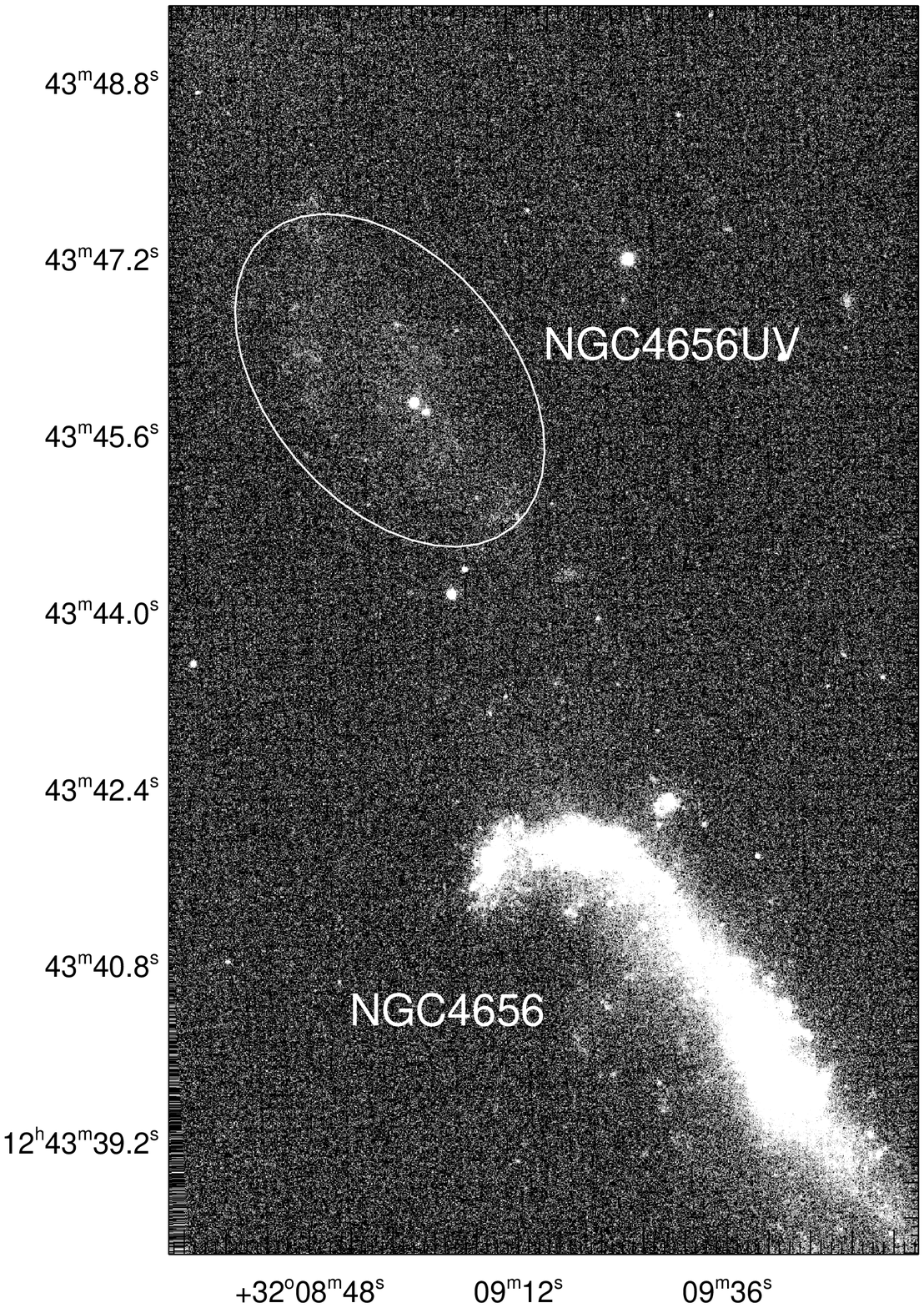}  
\hspace{-3.5cm} 
\includegraphics[scale=0.32]{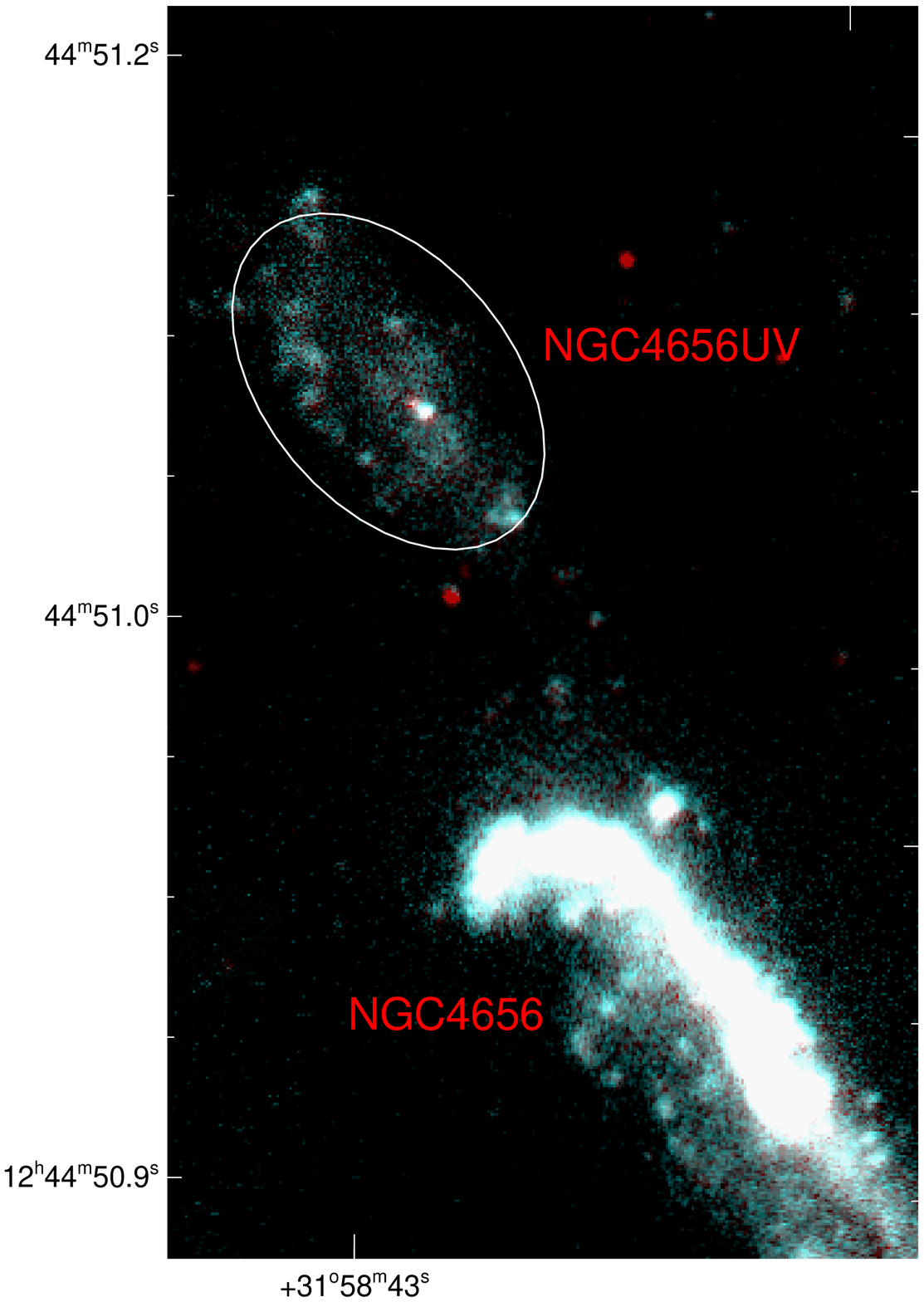}  
\hspace{-3.5cm}
\includegraphics[scale=0.32]{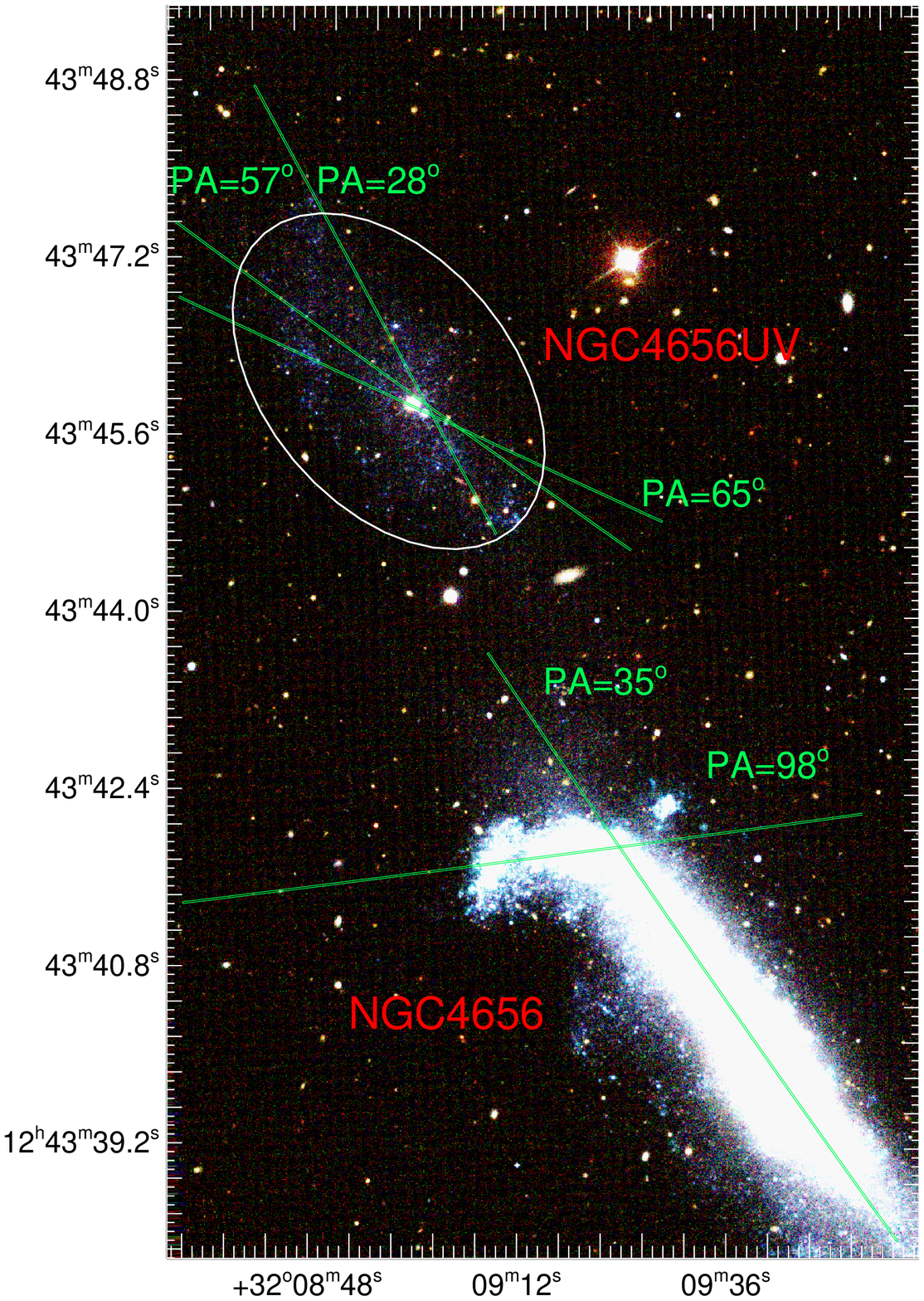}  
\vspace{-2cm}  
\caption{ The images of NGC4656 and NGC4656UV from left to right: in  u-band, FUV and NUV- bands and a gri- bands. The last one is shown with overplotted positions of the slit. The major axis of the ellipse overlaid in the images is 7 kpc.    }
\label{images} 
\end{figure*}

The second galaxy, NGC4656, is not so disturbed as its companion, although its surface brightness distribution is highly asymmetric, so that the NE-half contains the largest H~\textsc{ii} regions beyond the central area, and looks much brighter than the dim opposite half.  UV images of this galaxy obtained by GALEX allowed to reveal the  unusual satellite of about 10 kpc in  diameter which is  barely seen in the SDSS maps \citep{Schechtman-Rooketal2012} (see Fig. \ref{images}). This peculiar object -- NGC4656UV --- is a low surface brightness galaxy if to admit that it is a bound system. It  is connected with NGC4656 by \HI and probably stellar bridge. A  systemic velocity of NGC4656UV is close to that for NGC4656, which leaves no doubt that this object is a part of  the interacting system. Multiwavelength archival data analysis  carried out by \cite{Schechtman-Rooketal2012} for NGC4656, led the authors to the conclusions  that  NGC4656UV is a low-metallicity tidal dwarf candidate, containing  a  relatively low amounts of dark matter, although the other scenarios were also considered. 

The distance to NGC4656 and NGC4656UV is not well known. The brightest galaxies in the group -- NGC4656 and NGC4631 -- have systemic velocities with respect to Local Group  $V_{LG}$=  636 and 665 $\mathrm{km~s^{-1}}$ respectively (according to Hyperleda\footnote{http://leda.univ-lyon1.fr/}  database, \citep{Makarovetal2014}), corresponding to about 9 Mpc from the Hubble relation. The Updated Catalog of nearby galaxies \citep{Karachentsev2013}  gives the distance 5.4 Mpc for NGC4656  (TF-method), although the direct distance estimate for NGC4631 (TRGB-method)  gives 7.38 Mpc.  \cite{Schechtman-Rooketal2012} used the distance 7.2 Mpc, following \cite{Sethetal2005} (TRGB-method for NGC4656). We adopted this value for the system. 

 An ultra-faint $H\alpha$ "sheet" has been revealed between NGC 4656 and NGC 4631 by \cite{Donahue1995}. However, as it was noted in \cite{Schechtman-Rooketal2012}, NGC 4656UV is not detected in these observations.


\begin{figure}                                                               
 \centering                                                                   
 \includegraphics[width=\linewidth]{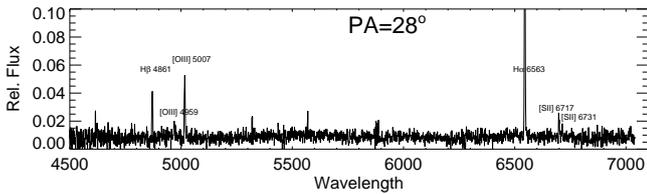}    
  \caption{ The sky-subtracted spectrum of NGC4656UV   for PA=28\degr, integrated along the slit. The flux is in units of $10^{-15} erg/cm^2/sec/$\AA, the wavelength is in \AA, (non-corrected for the redshift).  }
\label{spectra} 
\end{figure}

\begin{figure}                                                              
 \centering     
\hspace{-2.2cm} 
\includegraphics[scale=0.35]{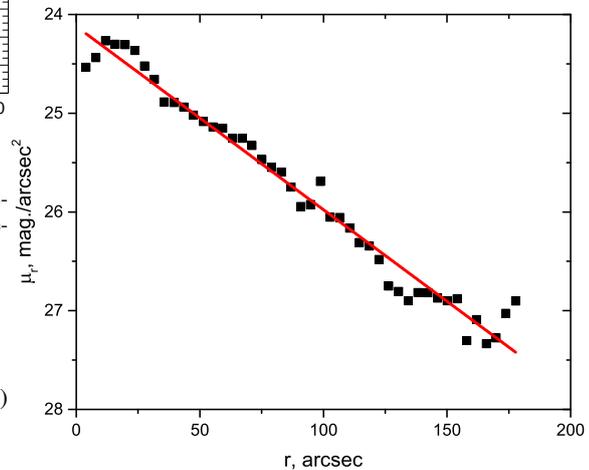}  
\vspace{-0.5cm} 
\caption{Radial profile of surface brightness of NGC4656UV obtained from SDSS-r data. The line corresponds to the exponential disc fit. No correction for disc inclination is applied.}
\label{profile_r} 
\end{figure}

The absence of spectral optical data makes difficult to clarify the nature of NGC4656UV. In this paper we describe the attempt to measure the velocity and metallicity of emission gas  of NGC4656UV and NGC4656 using the long-slit mode of observation at 6m telescope BTA of Special Astrophysical Observatory.  

The current paper is organized as follows: in Sect. \ref{Obs} we describe details and results of the data reduction, Sect.  \ref{Discussion} is devoted to the discussion and in Sect \ref{conclusion} we give the main conclusions.

\section{Long-slit observations}\label{Obs}
\subsection{Data reduction}
We observed NGC4656 and NGC4656UV in 2013-2016 with the spectrographs SCORPIO-2 \citep{AfanasievMoiseev2011} and SCORPIO \citep{AfanasievMoiseev2005}  at the
prime focus of the 6-m Russian telescope BTA at Special
Astrophysical Observatory of the Russian Academy of Sciences (SAO RAS). The dates of observations, exposure times, seeing, observers and dispersers are given in Table \ref{log}. In our observations we used two different dispersers: the grism VPHG1200@540 which covers the spectral range 3600-7070 \AA\, and has a dispersion of 0.87 \AA ~pixel$^{-1}$, spectral resolution $FWHM\approx 5.2$ \AA\,  and the grism VPHG2300G with the spectral range of 4800-5570 \AA, reciprocal dispersion 0.38 \AA/px and the spectral resolution of 2.2 \AA. The scale along the slit is 0.36 \arcsec/px, the slit width is 1\arcsec.The positions of the slit are shown in Fig. \ref{images} (right panel). 

The data reduction was performed in a standard way using   \textsc{idl}-based reduction pipeline. The full description of the reduction can be found e.g. in \cite{zasovetal2016} and \cite{zasovetal2015}. Shortly, the procedure included a bias subtraction and truncation of overscan regions; a division by normalized flat field frames; the wavelength calibration using the spectrum of a He-Ne-Ar calibration lamp; linearization and summation; the night sky subtraction; the flux calibration using the spectra of the standards HZ44 and Feige 56 that were obtained during the same or adjacent nights. The variation of instrumental profile of the spectrograph was measured and taken into account.  The sky-subtracted spectrum of NGC4656UV   for PA=28\degr, integrated along the slit is demonstrated in Fig. \ref{spectra}.

We fitted the reduced and binned spectra with high-resolution PEGASE.HR simple stellar population models \citep{LeBorgne2004} convolved with the variation of the instrumental profile using \textsc{NBursts} full spectral fitting technique  \citep{Chilingarian2007} (for more details see e.g. \citealt{zasovetal2016}). 
From this fitting we get the  parameters of stellar population:  age and metallicity, line-of-sight velocity, velocity dispersion and Gauss-Hermite moments h3 and h4 which characterize the deviation of LOSVD from the Gaussian profile. In order to obtain the velocity and velocity dispersion of the ionized gas as well as the fluxes of emission lines  we subtracted the stellar population model spectra from the observed ones and fitted the emission lines by Gaussian distribution.

The oxygen abundance, which is the indicator of gas metallicity, was estimated using the measured fluxes of emission lines. We were able to use `direct' $T_e$ method only for several brightest regions of NGC4656 because of the faintness of sensitive to electron temperature [O~\textsc{iii}] 4363 \AA\, emission line. Therefore, we used several calibrations that are based on the strong emission lines. There is a large discrepancy (up to 0.5 dex) between the metallicity estimates obtained by different methods \citep[see, e.g.,][]{kewley08, lopez-sanchez12}. We selected one empirical (calibrated by the sample of H~\textsc{ii} regions with well measured metallicity using $T_e$ method) and one theoretical (calibrated using photoionization models with pre-defined metallicity) methods: S method proposed by \cite{pilyugin16} and $izi$ method from \cite{izi}, respectively. It is necessary to know the relative ratio of [O~\textsc{iii}] 5007 \AA, [N~\textsc{ii}] 6584 \AA\, and [S~\textsc{ii}] 6717,6731 \AA\, to H$\beta$ fluxes in order to apply S-method, while $izi$ uses Bayesian inference to estimate oxygen abundance from all available emission lines fluxes. Note that we do not utilize the R-method proposed by Pilyugin \& Grebel, because it uses [O~\textsc{ii}] 3727 \AA\, line, which is very noisy for significant part of our spectra (yet there is a good agreement between the estimates obtained by S and R methods for the area of bright [O~\textsc{ii}] line). We will refer further to the oxygen abundance estimated using S and $izi$ methods as $12+\log\mathrm{(O/H)_S}$ and $12+\log\mathrm{(O/H)_{izi}}$ respectively.

\begin{table*}
\caption{Log of observations}\label{log}
\begin{center}
\begin{tabular}{cccccc}
\hline\hline
Slit PA & Date & Exposure time& Seeing& Observers & Disperser \\
   (\degr)  &  &     (s) &        (\arcsec)&& \\
\hline
\multicolumn{6}{c}{NGC4656}\\
\hline
35&08.02.2013&1800&1.5&Uklein, Katkov& VPHG1200@540 \\
98&08.02.2013&2700&1.5&Uklein, Katkov& VPHG1200@540 \\
\hline
\multicolumn{6}{c}{NGC4656UV}\\
\hline
28&31.03.2016&5400& 1.7& Uklein& VPHG1200@540 \\ 
57&06.04.2016&2700& 1.2& Uklein& VPHG1200@540 \\ 
65&09.05.2016&8400& 1.9& Uklein& VPHG2300G \\ 
\hline\hline
\end{tabular}
\end{center}
\end{table*}

\subsection{The estimates of kinematical parameters and ionized gas metallicity}
\subsubsection{NGC4656UV}
The enhanced  UV brightness of NGC4656UV and numerous  patches  observed mostly at its peripheric regions clearly demonstrate  the presence of young stellar population. The structure of the dwarf is rather uneven. However, the r-band surface brightness radial profile of NGC4656UV, that we obtained from SDSS data, follows the distribution law of exponential disc: \begin{equation}\mu_r(r)=(\mu_{0})_{r}+ 1.086(r/h), \end{equation} where $(\mu_{0})_{r}$ and $h$ are the disc central surface brightness and the exponential scalelength, correspondingly (see Fig. \ref{profile_r}).  The line in the figure denotes the result of the fitting with $(\mu_{0})_{r}= 24.12 ~\mathrm{mag\  arcsec^{-2}}$ (non-corrected for disc inclination) and $h=58\arcsec = 2.0$ kpc. It corresponds to the total luminosity $\sim 1.8\cdot10^8 L_\odot$, which after correction for the axes ratio of the outer isophotes $a/b\approx$ 2 gives $L_{r,c}\sim 0.9\cdot10^8 L_\odot$.

\begin{figure}                                                              
 \centering     
\includegraphics[width=\linewidth]{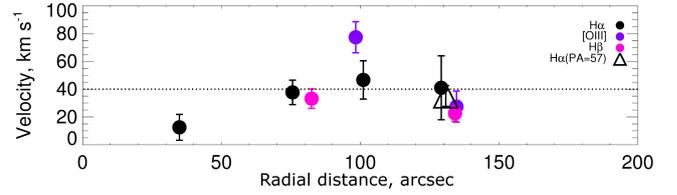}  
\caption{ The rotation velocities of ionized gas of NGC4656 obtained from two spectral slices PA=28\degr and PA=57\degr.  The weighted average value of rotation velocity from the approaching and receding sides is given for the large bins.  Dotted line shows the rotation velocity amplitude of 40 $\textrm{km~s$^{-1}$}$. }
\label{n4656rc} 
\end{figure}

A profile of rotational velocity of NGC4656UV is shown in Fig.\ref{n4656rc}. The points in the figure correspond to the weighted mean velocities in the radial bins of $\pm 20$ arcsec width.  Zeroth radial coordinate was chosen as the centre of symmetry of r-image, which is about 5 arcsec towards SE from the intersection point of the slits (see Fig.\ref{images}). Emission lines with low S/N ratio were ignored. The rotation curve was obtained using the emission spectra for two slices: PA=28\degr and PA=57\degr \footnote{ The positions of the slit are shown in Fig. \ref{images}, right panel.} and the following equations:
\begin{equation}
\label{formula_v} V(r) = \frac{V_r(r)\sqrt{(sec^2(i)-\tan^2(i)\cos^2(\phi))}}{\sin(i) \cos(\phi)}
\end{equation}

\begin{equation}
\label{formula_r} r = r_\phi \sqrt{(\sec^2(i) - \tan^2(i) \cos^2 (\alpha))}
\end{equation}
Here $r$ is the radial coordinate, $V(r)$ is circular velocity and $\phi$  is the angle between  radius-vector of a given point and the major axis of a galaxy,  $ r_\phi $ is the radial distance in the sky plane, $V_r$ is the line-of-sight velocity corrected for the systemic velocity (the approaching side is mirrored and averaged with the receding side) and $i$ is inclination angle, which corresponds to  the observed isophotes ratio a/b $\sim $2. The systemic velocity of the galaxy is taken to be 590 $\mathrm{km~s^{-1}}$. 

The maximum velocity of rotation $V=40\pm 10$ $\mathrm{km~s^{-1}}$ allows us to get the rough estimate of total dynamical mass of the galaxy inside of radius r=3h=174\arcsec= 6 kpc: $M \approx V ^2\cdot r/G \approx 2.2 (+ 1.2, -1) \cdot10^{9} M_\odot$. This value is in good agreement with the result of \cite{Schechtman-Rooketal2012} obtained from the PV diagram of \HI. It corresponds to the dynamical mass-to-light ratio $M_{dyn}/L_r\sim 10-30 M_\odot / L_\odot $. A blue colour of this galaxy (according to \citealt{Schechtman-Rooketal2012}, $(g-r) \approx 0$) corresponds to $M_*/L_r \approx$ 0.16-0.25$ M_\odot / L_\odot $ for different stellar population models \citep[see f.e.][]{IntoPortinari2013, RoedigerCourteau2015}. Hence the input of stellar mass into the total mass of the dwarf is low. The mass of gas prevails over the mass of stars:  \HI mass of NGC 4656UV (estimated as $3.8 \cdot10^{8} M_\odot $  by \cite{Schechtman-Rooketal2012}), corresponds to the total mass of gas (including helium) $\sim5\cdot 10^8 M_\odot$. Even if we assume that most of the observed gas connected with this galaxy lays within the optical borders of NGC 4656UV, although this is a clear exaggeration,  and keep in mind that the gas mass is underestimated due to the unaccounted mass of the molecular gas (which is expected to be quite low since the metallicity of the gas is low), the  conclusion is unavoidable that the observed baryonic mass is several times lower than the dynamic mass in this galaxy. Thus it is unlikely that NGC4656UV is a tidal dwarf,  which should contain no dark matter (unless one supposes that NGC 4656UV has  large amount of dark gas (i.e. cold gas non-detected by its emission, see, e.g. \citealt{Kasparova2014})). Another argument against the tidal formation of the galaxy is its large size which is only 2-3 times lower than that of NGC 4656 parallel with the exponential shape of its surface brightness radial profile which evidences that NGC 4656UV is rather relaxed system.  Most probably we have a deal with the dark matter dominated LSB dwarf galaxy.  One should keep in mind, however, that the total dynamical mass estimate of NGC4656UV remains uncertain and needs more elaborate investigation.

Because of the weakness of even strong emission lines in the spectra of NGC 4656UV, we were able to measure the oxygen abundance for only one H~\textsc{ii} region, which reveals the brightest emission in [O~\textsc{iii}] 5007 \AA,  [N~\textsc{ii}] 6584 \AA\, and [S~\textsc{ii}] 6717,6731 \AA\, among all studied areas in this galaxy. This region is located at the south-west edge of the galaxy and crossed by the slit with PA=28\degr (see Fig.~\ref{images}). We got the following oxygen abundances estimates for it: $12+\log\mathrm{(O/H)_S}=7.28 \pm 0.10$ and $12+\log\mathrm{(O/H)_{izi}}=7.85 \pm 0.12$. As we already noted in Section~\ref{Obs}, the discrepancy between these estimates is a very common problem: theoretical methods (like $izi$) usually yield values higher (up to 0.5~dex) than empirical ones. Our estimates of oxygen abundance in NGC4656UV confirm the findings of  \cite{Schechtman-Rooketal2012}, who supposed that it should contain very low metallicity gas in order to explain the lack of IR emission.

	

\subsubsection{NGC4656}
\begin{figure} 
	\includegraphics[width=\linewidth]{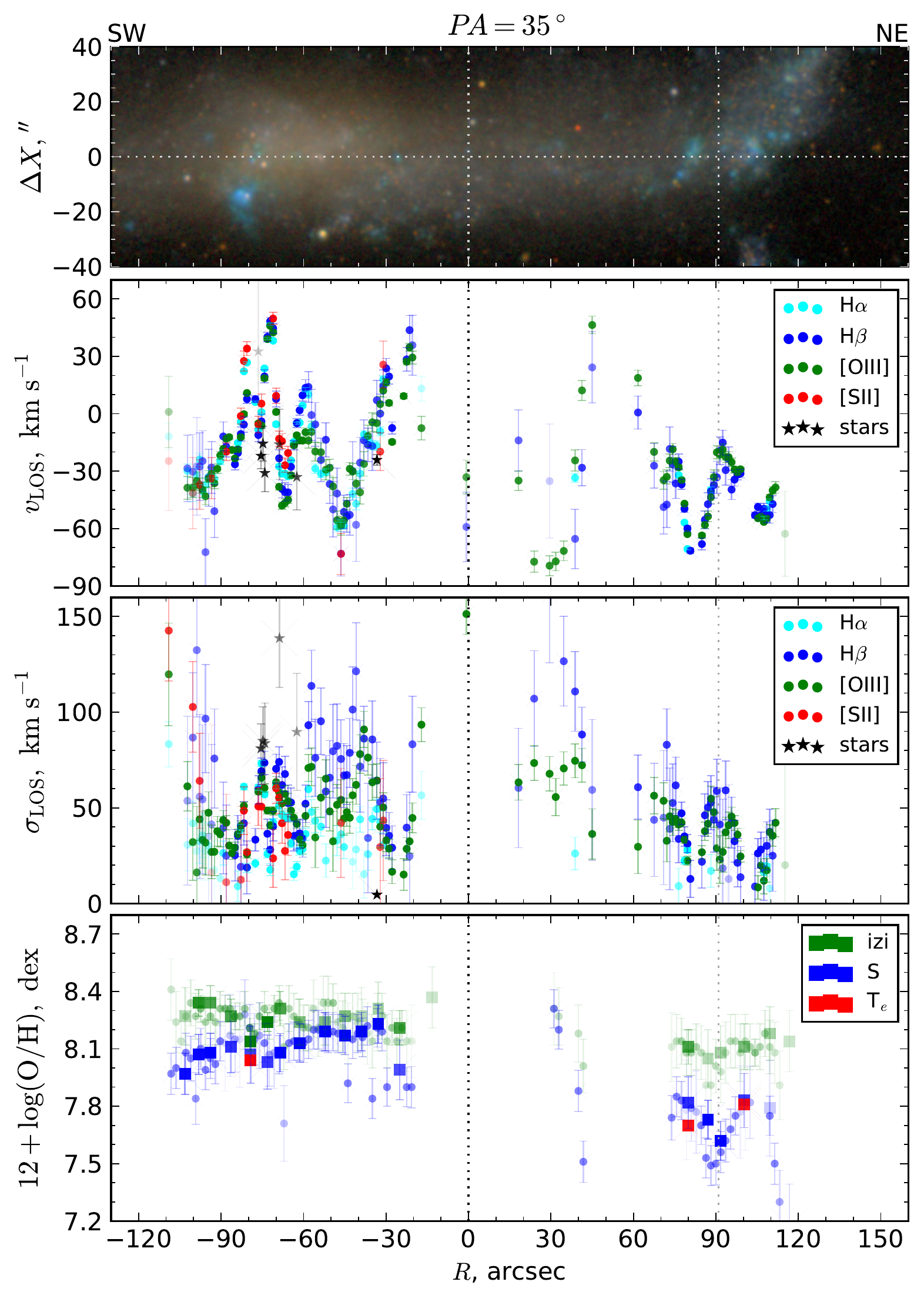} 
	\caption{
		Radial distribution of line-of-sight velocity, velocity dispersion and oxygen abundance of NGC4656 along the slit PA=35\degr. Top panel corresponds to the reference SDSS {\it gri} composite image. The zero point for the velocity $V_{sys}=646~\mathrm{km~s^{-1}}$. Circles at the bottom panel correspond to the values obtained for each bin along the slit, while squares represent the values measured by integrated spectra of H~\textsc{ii} regions listed in Table~\ref{tab:spec}. Zero-point for radial distance is arbitrary chosen.}
	
	\label{kin_prof_215}
\end{figure}

\begin{figure} 
	\includegraphics[width=\linewidth]{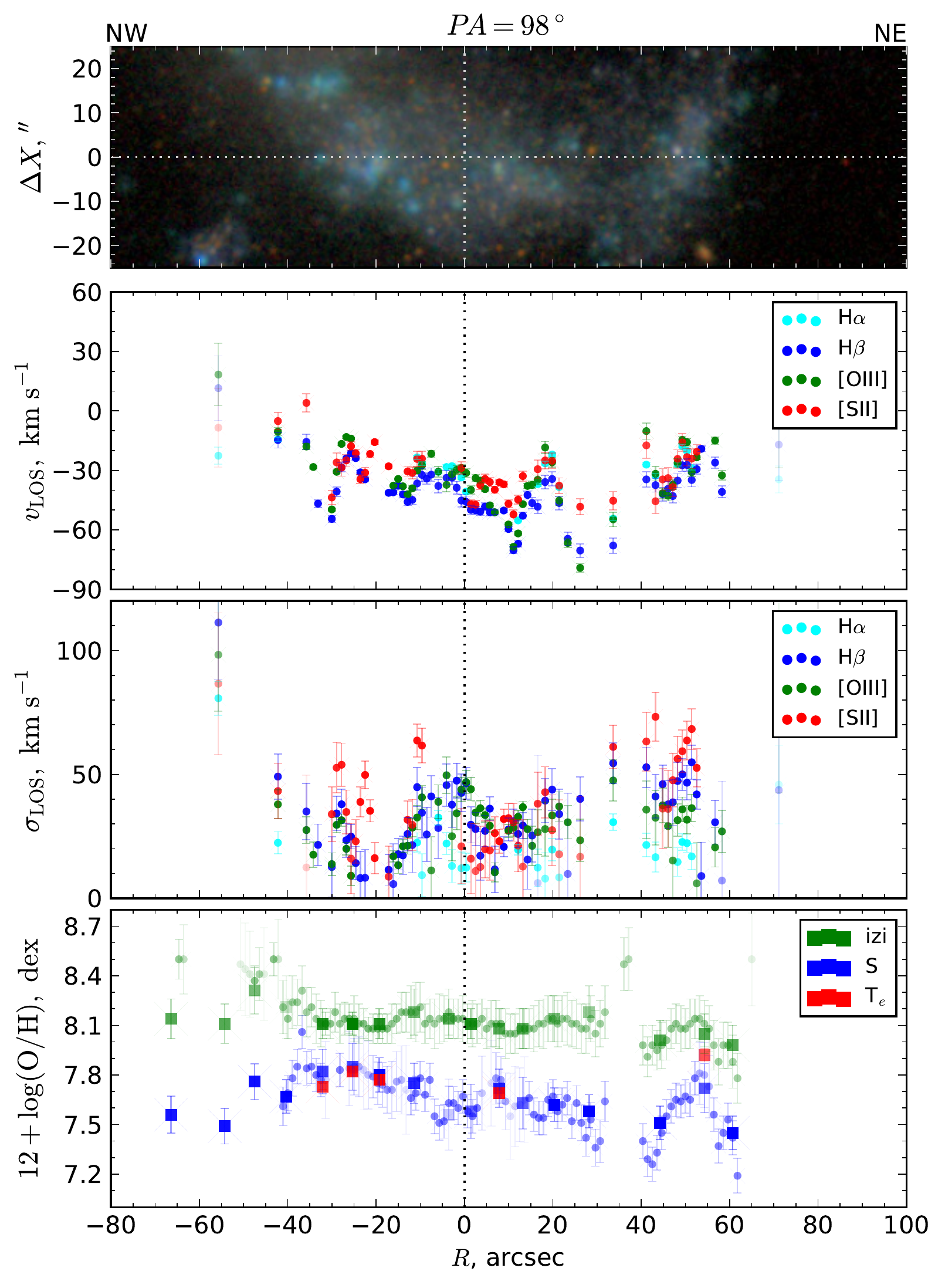}
	\caption{The same as in Fig. \ref{kin_prof_215}, but  for PA=98\degr.} 
	\label{kin_prof_278}
\end{figure}

In Figs. \ref{kin_prof_215}, \ref{kin_prof_278} we demonstrate the radial profiles of the velocity and velocity dispersion for PA=35\degr and PA=98\degr respectively. The circles correspond to the ionized gas, asterisks denote the estimates for stellar population. For  PA=98\degr the parameters of stars appeared to be very uncertain, thus we decided not to put them into the graphs. 

\begin{landscape}
	\begin{table}
		\caption{Results of spectroscopy of H~\textsc{ii} regions in NGC4656UV and NGC4656 galaxies}\label{tab:spec}
		\tiny
		\begin{tabular}{ccccccccccccccc}
			\hline
			Reg. & Pos., & F(H$\beta$),  & [O~\textsc{ii}]3727 & H$\gamma$ & [O~\textsc{iii}]4363 & [O~\textsc{iii}]5007 & H$\alpha$ & [N~\textsc{ii}]6584 & [S~\textsc{ii}]6717 & [S~\textsc{ii}]6731 & c(H$\beta$) & $12+\log\mathrm{(O/H)_S}$ & $12+\log\mathrm{(O/H)_{izi}}$ & $12+\log\mathrm{(O/H)_{Te}}$ \\
			& arcsec & $10^{-16}~\mathrm{erg s^{-1}cm{-2}}$ &  &  &  &  &  &  &  &  &  &  &  &  \\
			\hline
			\multicolumn{15}{c}{\small NGC4656UV PA28}\\
			\hline
			1 & $59\div62$ & $2.14\pm0.02$ & $-$ &$-$ &$-$ &$0.532\pm0.011$ & $2.862\pm0.430$ & $0.119\pm0.006$ & $0.223\pm0.005$ & $0.137\pm0.009$ & $0.47$ &$7.28\pm0.10$ &$7.85\pm0.12$ &$-$ \\
			\hline
			\multicolumn{15}{c}{\small NGC4656 PA35}\\
			\hline
			1 & $-104\div-101$ & $7.89\pm0.31$ & $-$ &$-$ &$-$ &$2.112\pm0.088$ & $2.698\pm1.117$ & $0.177\pm0.024$ & $0.394\pm0.023$ & $0.347\pm0.046$ & $0.00$ &$7.97\pm0.11$ &$8.27\pm0.15$ &$-$ \\
			2 & $-99\div-96$ & $13.43\pm0.39$ & $-$ &$-$ &$-$ &$3.295\pm0.098$ & $2.861\pm0.945$ & $0.153\pm0.017$ & $0.514\pm0.021$ & $0.333\pm0.028$ & $0.13$ &$8.07\pm0.10$ &$8.34\pm0.07$ &$-$ \\
			3 & $-95\div-92$ & $17.56\pm0.36$ & $-$ &$0.720\pm0.063$ & $-$ &$3.362\pm0.071$ & $2.861\pm0.500$ & $0.209\pm0.013$ & $0.467\pm0.015$ & $0.339\pm0.024$ & $0.14$ &$8.08\pm0.10$ &$8.34\pm0.09$ &$-$ \\
			4 & $-89\div-82$ & $200.89\pm0.77$ & $3.093\pm0.072$ & $0.418\pm0.009$ & $-$ &$3.852\pm0.015$ & $2.861\pm0.102$ & $0.154\pm0.002$ & $0.338\pm0.002$ & $0.236\pm0.004$ & $0.30$ &$8.11\pm0.13$ &$8.27\pm0.09$ &$-$ \\
			5 & $-81\div-77$ & $728.79\pm1.16$ & $1.857\pm0.033$ & $0.398\pm0.004$ & $0.092\pm0.010$ & $6.323\pm0.010$ & $2.862\pm0.080$ & $0.071\pm0.001$ & $0.121\pm0.001$ & $0.098\pm0.002$ & $0.55$ &$8.07\pm0.15$ &$8.14\pm0.05$ &$8.04\pm0.03$ \\
			6 & $-75\div-70$ & $210.91\pm0.72$ & $3.486\pm0.064$ & $0.362\pm0.008$ & $-$ &$3.218\pm0.011$ & $2.861\pm0.103$ & $0.151\pm0.002$ & $0.281\pm0.002$ & $0.198\pm0.004$ & $0.22$ &$8.03\pm0.14$ &$8.24\pm0.05$ &$-$ \\
			7 & $-70\div-66$ & $83.49\pm0.55$ & $5.384\pm0.127$ & $0.424\pm0.014$ & $-$ &$2.400\pm0.017$ & $2.861\pm0.143$ & $0.203\pm0.004$ & $0.431\pm0.005$ & $0.316\pm0.008$ & $0.23$ &$8.08\pm0.10$ &$8.31\pm0.09$ &$-$ \\
			8 & $-65\div-57$ & $259.67\pm0.82$ & $3.655\pm0.068$ & $0.410\pm0.006$ & $-$ &$2.108\pm0.007$ & $2.861\pm0.067$ & $0.196\pm0.002$ & $0.331\pm0.002$ & $0.236\pm0.003$ & $0.24$ &$8.13\pm0.10$ &$8.24\pm0.12$ &$-$ \\
			9 & $-55\div-48$ & $35.53\pm0.60$ & $-$ &$0.308\pm0.016$ & $-$ &$1.410\pm0.026$ & $2.518\pm0.213$ & $0.263\pm0.009$ & $0.394\pm0.009$ & $0.280\pm0.013$ & $0.00$ &$8.19\pm0.10$ &$8.24\pm0.12$ &$-$ \\
			10 & $-47\div-42$ & $26.61\pm0.36$ & $4.239\pm0.233$ & $0.364\pm0.018$ & $-$ &$2.050\pm0.029$ & $2.812\pm0.257$ & $0.240\pm0.008$ & $0.358\pm0.010$ & $0.256\pm0.017$ & $0.00$ &$8.17\pm0.10$ &$8.27\pm0.12$ &$-$ \\
			11 & $-41\div-36$ & $19.35\pm0.37$ & $-$ &$0.520\pm0.033$ & $-$ &$1.725\pm0.037$ & $2.819\pm0.326$ & $0.272\pm0.012$ & $0.401\pm0.012$ & $0.263\pm0.017$ & $0.00$ &$8.19\pm0.10$ &$8.27\pm0.15$ &$-$ \\
			12 & $-34\div-30$ & $25.31\pm0.41$ & $-$ &$0.563\pm0.034$ & $-$ &$2.154\pm0.038$ & $2.861\pm0.300$ & $0.284\pm0.011$ & $0.323\pm0.010$ & $0.250\pm0.018$ & $0.18$ &$8.23\pm0.10$ &$8.31\pm0.14$ &$-$ \\
			13 & $-30\div-19$ & $172.17\pm0.49$ & $3.381\pm0.058$ & $0.455\pm0.007$ & $-$ &$2.776\pm0.008$ & $2.860\pm0.083$ & $0.158\pm0.002$ & $0.209\pm0.002$ & $0.150\pm0.003$ & $0.06$ &$7.99\pm0.16$ &$8.21\pm0.09$ &$-$ \\
			14 & $-17\div-9$ & $8.27\pm0.31$ & $-$ &$-$ &$-$ &$4.317\pm0.181$ & $2.861\pm1.383$ & $0.254\pm0.040$ & $-$ &$-$ &$0.23$ &$-$ &$8.37\pm0.16$ &$-$ \\
			15 & $75\div84$ & $343.12\pm0.49$ & $2.056\pm0.036$ & $0.462\pm0.003$ & $0.051\pm0.010$ & $3.052\pm0.005$ & $2.860\pm0.083$ & $0.083\pm0.001$ & $0.185\pm0.001$ & $0.130\pm0.002$ & $0.10$ &$7.82\pm0.13$ &$8.11\pm0.09$ &$7.70\pm0.06$ \\
			16 & $85\div88$ & $24.72\pm0.23$ & $-$ &$0.492\pm0.020$ & $-$ &$1.925\pm0.019$ & $2.567\pm0.431$ & $0.097\pm0.006$ & $0.210\pm0.005$ & $0.150\pm0.009$ & $0.00$ &$7.73\pm0.10$ &$8.05\pm0.15$ &$-$ \\
			17 & $89\div94$ & $30.34\pm0.29$ & $-$ &$0.417\pm0.023$ & $-$ &$1.844\pm0.019$ & $2.860\pm0.626$ & $0.080\pm0.006$ & $0.243\pm0.006$ & $0.152\pm0.009$ & $0.01$ &$7.62\pm0.10$ &$8.08\pm0.14$ &$-$ \\
			18 & $95\div104$ & $380.95\pm0.58$ & $1.867\pm0.041$ & $0.470\pm0.004$ & $0.054\pm0.011$ & $3.584\pm0.006$ & $2.861\pm0.097$ & $0.070\pm0.001$ & $0.172\pm0.001$ & $0.121\pm0.002$ & $0.17$ &$7.83\pm0.14$ &$8.11\pm0.12$ &$7.81\pm0.06$ \\
			19 & $106\div112$ & $65.03\pm0.29$ & $3.250\pm0.125$ & $0.469\pm0.012$ & $-$ &$2.713\pm0.013$ & $2.684\pm0.281$ & $0.081\pm0.003$ & $0.242\pm0.003$ & $0.169\pm0.005$ & $0.00$ &$7.79\pm0.25$ &$8.18\pm0.14$ &$-$ \\
			20 & $114\div118$ & $8.45\pm0.25$ & $-$ &$-$ &$-$ &$0.858\pm0.036$ & $2.412\pm5.132$ & $0.033\pm0.019$ & $0.461\pm0.020$ & $0.272\pm0.027$ & $0.00$ &$7.17\pm0.22$ &$8.14\pm0.16$ &$-$ \\
			\hline
			\multicolumn{15}{c}{\small NGC4656 PA98}\\
			\hline
			21 & $-69\div-63$ & $1.95\pm0.06$ & $-$ &$-$ &$-$ &$0.660\pm0.029$ & $2.721\pm1.154$ & $0.167\pm0.023$ & $0.542\pm0.024$ & $0.325\pm0.032$ & $0.00$ &$7.56\pm0.11$ &$8.14\pm0.12$ &$-$ \\
			22 & $-56\div-51$ & $2.19\pm0.06$ & $-$ &$-$ &$-$ &$0.618\pm0.026$ & $2.860\pm1.176$ & $0.157\pm0.021$ & $0.479\pm0.021$ & $0.328\pm0.034$ & $0.07$ &$7.49\pm0.11$ &$8.11\pm0.12$ &$-$ \\
			23 & $-50\div-44$ & $4.15\pm0.06$ & $-$ &$-$ &$-$ &$2.088\pm0.031$ & $2.860\pm0.982$ & $0.098\pm0.011$ & $0.417\pm0.011$ & $0.313\pm0.020$ & $0.09$ &$7.76\pm0.11$ &$8.31\pm0.14$ &$-$ \\
			24 & $-43\div-37$ & $7.12\pm0.08$ & $-$ &$-$ &$-$ &$1.290\pm0.017$ & $2.861\pm0.533$ & $0.124\pm0.008$ & $0.454\pm0.008$ & $0.348\pm0.015$ & $0.21$ &$7.67\pm0.10$ &$8.21\pm0.19$ &$-$ \\
			25 & $-34\div-29$ & $45.35\pm0.05$ & $2.749\pm0.039$ & $0.463\pm0.003$ & $0.050\pm0.007$ & $3.015\pm0.004$ & $2.560\pm0.084$ & $0.075\pm0.001$ & $0.194\pm0.001$ & $0.134\pm0.002$ & $0.00$ &$7.82\pm0.13$ &$8.11\pm0.09$ &$7.73\pm0.06$ \\
			26 & $-27\div-22$ & $120.97\pm0.11$ & $2.172\pm0.035$ & $0.457\pm0.002$ & $0.067\pm0.006$ & $4.087\pm0.004$ & $2.861\pm0.067$ & $0.062\pm0.001$ & $0.144\pm0.001$ & $0.102\pm0.001$ & $0.27$ &$7.85\pm0.14$ &$8.11\pm0.07$ &$7.82\pm0.03$ \\
			27 & $-22\div-16$ & $283.00\pm0.11$ & $1.815\pm0.012$ & $0.439\pm0.001$ & $0.057\pm0.002$ & $3.575\pm0.001$ & $2.860\pm0.031$ & $0.062\pm0.001$ & $0.155\pm0.000$ & $0.111\pm0.001$ & $0.13$ &$7.80\pm0.10$ &$8.11\pm0.09$ &$7.77\pm0.001$ \\
			28 & $-14\div-7$ & $49.56\pm0.07$ & $2.778\pm0.050$ & $0.374\pm0.003$ & $-$ &$2.170\pm0.003$ & $2.860\pm0.083$ & $0.098\pm0.001$ & $0.315\pm0.001$ & $0.215\pm0.002$ & $0.05$ &$7.75\pm0.12$ &$8.18\pm0.14$ &$-$ \\
			29 & $-5\div-1$ & $17.70\pm0.07$ & $2.868\pm0.171$ & $-$ &$-$ &$1.058\pm0.005$ & $2.861\pm0.191$ & $0.113\pm0.003$ & $0.435\pm0.003$ & $0.292\pm0.005$ & $0.17$ &$7.56\pm0.21$ &$8.14\pm0.12$ &$-$ \\
			30 & $0\div3$ & $33.92\pm0.07$ & $3.509\pm0.100$ & $0.460\pm0.005$ & $-$ &$1.109\pm0.003$ & $2.861\pm0.092$ & $0.117\pm0.001$ & $0.359\pm0.001$ & $0.249\pm0.002$ & $0.18$ &$7.58\pm0.16$ &$8.11\pm0.09$ &$-$ \\
			31 & $5\div10$ & $137.16\pm0.11$ & $2.246\pm0.031$ & $0.448\pm0.002$ & $0.040\pm0.004$ & $2.525\pm0.002$ & $2.861\pm0.048$ & $0.076\pm0.001$ & $0.212\pm0.001$ & $0.151\pm0.001$ & $0.28$ &$7.72\pm0.12$ &$8.08\pm0.12$ &$7.69\pm0.03$ \\
			32 & $11\div15$ & $41.61\pm0.06$ & $3.277\pm0.065$ & $0.439\pm0.004$ & $-$ &$1.531\pm0.003$ & $2.860\pm0.082$ & $0.101\pm0.001$ & $0.250\pm0.001$ & $0.178\pm0.002$ & $0.10$ &$7.63\pm0.16$ &$8.08\pm0.12$ &$-$ \\
			33 & $17\div23$ & $17.05\pm0.06$ & $-$ &$0.411\pm0.008$ & $-$ &$1.316\pm0.006$ & $2.860\pm0.201$ & $0.110\pm0.003$ & $0.375\pm0.003$ & $0.253\pm0.005$ & $0.13$ &$7.62\pm0.10$ &$8.14\pm0.14$ &$-$ \\
			34 & $24\div32$ & $9.10\pm0.06$ & $-$ &$0.473\pm0.017$ & $-$ &$1.163\pm0.009$ & $2.855\pm0.378$ & $0.106\pm0.005$ & $0.456\pm0.005$ & $0.315\pm0.009$ & $0.00$ &$7.58\pm0.10$ &$8.18\pm0.16$ &$-$ \\
			35 & $38\div49$ & $39.60\pm0.10$ & $-$ &$0.436\pm0.007$ & $-$ &$1.088\pm0.004$ & $2.861\pm0.149$ & $0.100\pm0.002$ & $0.274\pm0.002$ & $0.200\pm0.003$ & $0.17$ &$7.51\pm0.10$ &$8.01\pm0.12$ &$-$ \\
			36 & $51\div56$ & $58.13\pm0.07$ & $2.498\pm0.052$ & $0.433\pm0.003$ & $0.021\pm0.005$ & $1.992\pm0.003$ & $2.860\pm0.064$ & $0.100\pm0.001$ & $0.213\pm0.001$ & $0.150\pm0.001$ & $0.11$ &$7.72\pm0.15$ &$8.05\pm0.12$ &$7.92\pm0.12$ \\
			37 & $57\div63$ & $14.03\pm0.04$ & $-$ &$0.510\pm0.008$ & $-$ &$1.074\pm0.004$ & $2.775\pm0.250$ & $0.084\pm0.003$ & $0.257\pm0.003$ & $0.186\pm0.005$ & $0.00$ &$7.45\pm0.10$ &$7.98\pm0.12$ &$-$ \\
			\hline
		\end{tabular}
	\end{table}
\end{landscape}

Velocity profile along PA=35\degr reveals a velocity gradient along the galaxy, evidently reflecting the disc rotation and  large non-circular motions and unusually high velocity dispersion locally exceeding 100 $\textrm{km~s$^{-1}$}$. Non-circular gas motions are also  clearly seen along  PA=98\degr, passing through the NE part of galaxy, although the velocity wiggles are not so large as in the central  H~\textsc{ii} regions crossed by PA=98\degr. 
The restricted radial extension of our measurements for NGC4656 and non-circular velocities of emission gas we observe do not allow us to obtain the reliable value of rotational velocity amplitude for the galaxy. However, \HI data which are much more extended  enabled to determine the rotational velocity of $\sim 65$ $\textrm{km~s$^{-1}$}$ \cite{Schechtman-Rooketal2012}. It corresponds to the total dynamical mass of NGC4656 inside of $6-7$ kpc (the distance from the centre to the outer regions of the galaxy adjacent to UV dwarf) $M_{tot}\approx 7 \cdot10^{9} M_\odot$. 

In  Figs. \ref{kin_prof_215} and \ref{kin_prof_278}  we show also the distribution of oxygen abundance along the slits.
Parallel with the  oxygen abundance distribution for each pixel along the slit (shown by circles), we give the mean values obtained from the integrated spectra of individual H~\textsc{ii} regions or areas of diffuse emission (shown by squares). The results of the emission lines measuring and the metallicity estimates obtained for these selected regions are listed in Table~\ref{tab:spec}.

\begin{figure} 
	\includegraphics[width=\linewidth]{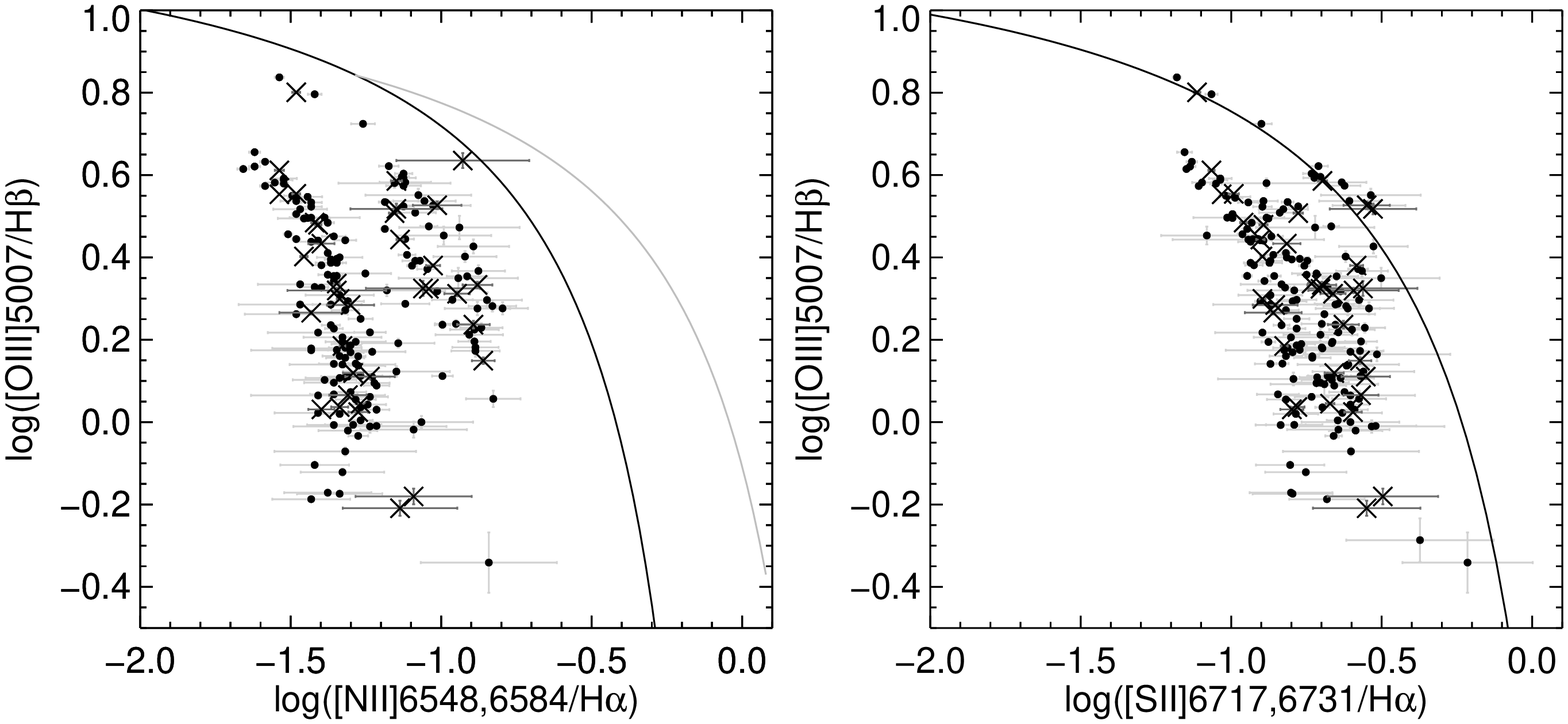}
	\caption{Diagnostic BPT diagrams [O~\textsc{iii}]/H$\beta$ vs [S~\textsc{ii}]/H$\alpha$ and [N~\textsc{ii}]/H$\alpha$ constructed for each pixel along the slits (shown by circles) and for individual H~\textsc{ii} regions from Table~\ref{tab:spec} (shown by crosses). Solid lines, according to the models of \citet{kewley2001} and \citet{kauffmann03}, separate regions of pure photoionization excitation caused by young massive stars (below the black line), shock excitation (or another mechanism not caused by star formation; above both black and grey lines) and combined contribution of both mechanisms (between grey and black lines in the lefthand panel).}
	\label{fig:bpt}
\end{figure}

In Fig.~\ref{fig:bpt} we show the so-called BPT diagnostic diagrams \citep{BPT} of emission lines flux ratios constructed for both slits. All studied regions fall in the area corresponding to the photoionization mechanism of excitation (under the black separation line from \cite{kewley2001}). It justifies the use of the  empirical methods to estimate oxygen abundance.

As it is seen from bottom panels of Figs. \ref{kin_prof_215}, \ref{kin_prof_278}, gas metallicity is higher at the southern (dimmest) part of NGC4656 galaxy and lower at the NE side of NGC4656, facing the UV satellite. For several brightest regions we were able to estimate oxygen abundance using $T_e$ method. The inferred values are in good agreement with estimates made with S method. 

\begin{figure} 
	\includegraphics[width=\linewidth]{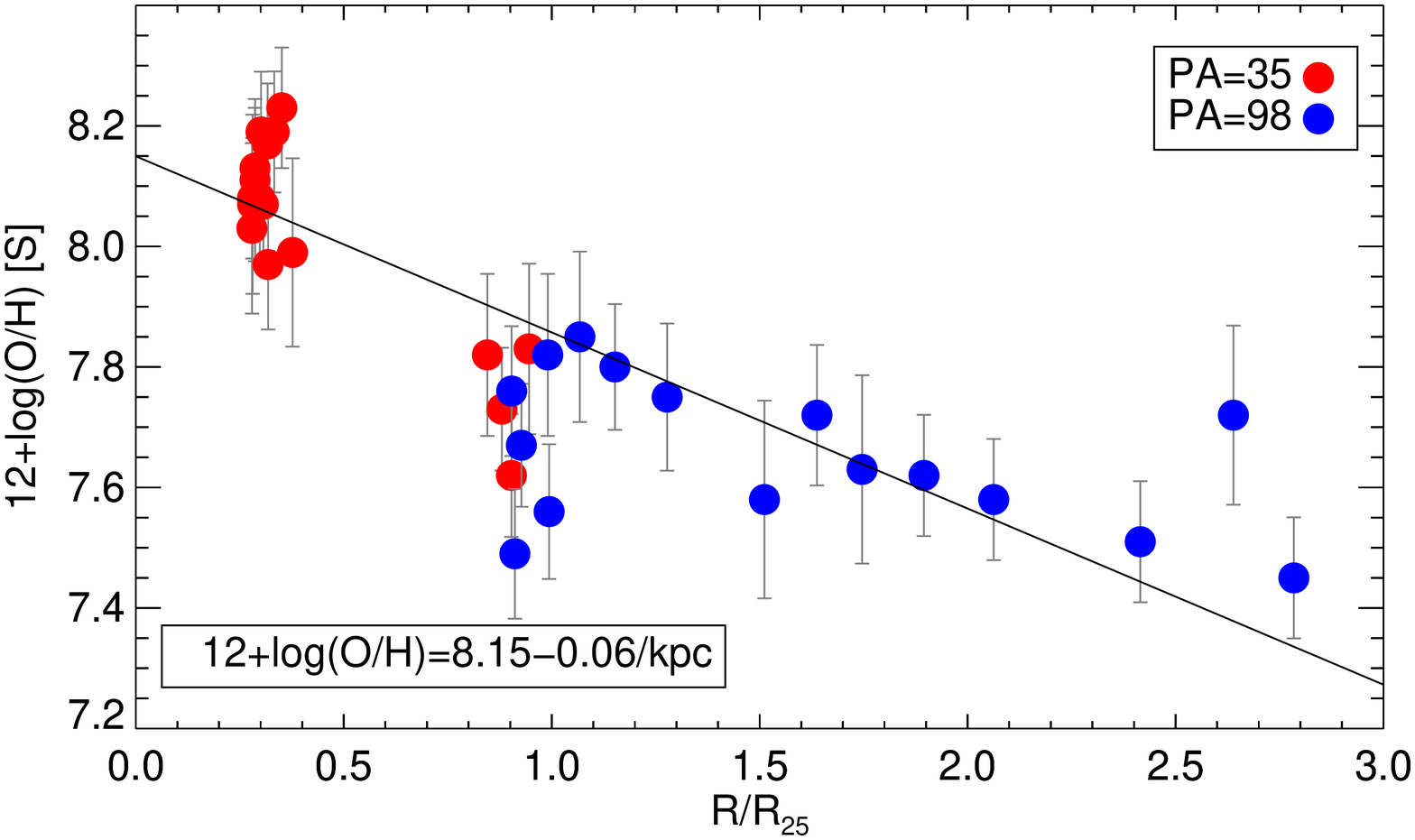}
	
	\includegraphics[width=\linewidth]{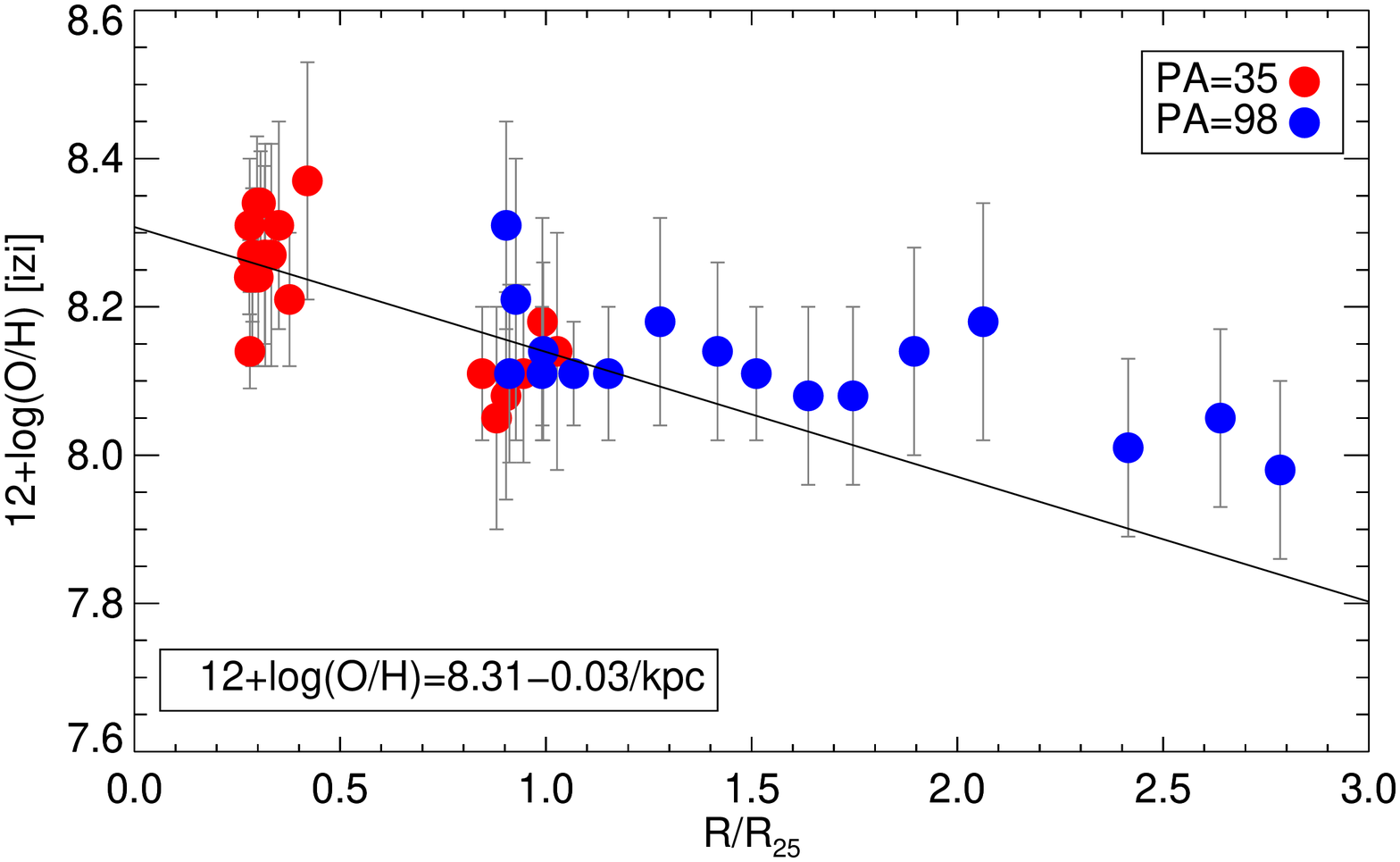}

	\caption{The variation of oxygen abundance with the galactocentric distance. The deprojected distance estimates are made using the morphological parameters of NGC 4656 (PA=35.6\degr and $R_{25}=4.93$ kpc) taken from HYPERLEDA database. The inclination of the galaxy was adopted as $i=79$\degr according to \citet{Schechtman-Rooketal2012}.} 
	\label{fig:abu_r}
\end{figure}

As it follows from Figs.~\ref{kin_prof_215} and \ref{kin_prof_278}, a gas is not chemically homogeneous in the NGC4656 galaxy. In Fig.~\ref{fig:abu_r} we show the oxygen abundance distribution along the deprojected galactocentric distance normalized to $R_{25} = 4.9$ kpc. To obtain the deprojected distance for each point, we adopted the following position angle and inclination of NGC4656: PA=35.6\degr (HYPERLEDA database), $i=79$\degr \citealt{Schechtman-Rooketal2012} respectively.  In Fig.~\ref{fig:abu_r} one may clearly see oxygen abundance gradient in NGC4656. This gradient found by izi-method (see Fig.~\ref{fig:abu_r}, below) is in a good agreement with the estimate of the O/H gradient from  \cite{Pilyuginetal2014} (grad O/H = 0.03 dex/kpc), who used the bright emission lines method.

An intriguing feature is seen at the northern part of the galaxy, facing the UV dwarf ($\sim 90$ arcsec along the slit PA=35, see Fig.\ref{kin_prof_215}). Both methods used for metallicity estimation show a drop of metallicity there: from  $12+\log\mathrm{(O/H)_S}=7.83\pm0.13$ and $12+\log\mathrm{(O/H)_{izi}}=8.18\pm0.14$ to $12+\log\mathrm{(O/H)_S}=7.63\pm0.09$ and $12+\log\mathrm{(O/H)_{izi}}=8.05\pm0.08$, which approaches  to the metallicity of UV dwarf ($12+\log\mathrm{(O/H)_{izi}}= 7.85$). It could be a consequence of the accretion  of metal-poor gas both at the UV dwarf and at the adjacent side of the main galaxy, which has inspired the intense star formation there. This hypothesis is consistent with the line-of-sight velocity distribution along the slit crossing NGC4656 along its major axis: it reveals the `V'-shape detail, where LOS velocity falls at $30-60 \textrm{km~s$^{-1}$}$, which also could be caused by the falling external gas. Note, however, that this kinematic feature is slightly shifted from the region of metallicity' minimum. Noteworthy that this site of possible gas accretion coincides with the region of the visual `bend' of galactic body.

\section{Discussion}\label{Discussion}

Two galaxies - NGC4656 and its UV satellite are in a process of tidal interaction. A mean density, which is  proportional to $M_{dyn}/R_{opt}^3$, for dwarf satellite is comparable with, if not higher than the density of the main galaxy, hence the satellite cannot be destroyed by tidal forces during the  close encounter of galaxies. Nevertheless the  tidal forces between these galaxies are responsible for formation of a short bridge connecting them.  
From \cite{Schechtman-Rooketal2012} (Fig 8) it follows that the line-of-sight velocity along the \HI- bridge changes at about 30 $\mathrm{km~s^{-1}}$  within the distance range of about 4\arcmin.  It corresponds to about 3.6 $\mathrm{km~s^{-1}}$/kpc for the assumed distance 7.2 Mpc. At the same time, our data give for the dwarf galaxy, which lies at the extension of the bridge, a velocity gradient ($30 \pm 5$ )  $\mathrm{km~s^{-1}}$ /100 \arcsec, or about ($8.5 \pm 1.5$) $\mathrm{km~s^{-1}}$/kpc. It definitely exceeds the velocity gradient along the bridge, which evidences that  the UV dwarf is not a part of the tidal feature coming out of the galaxy. We conclude that NGC4656UV is rather LSB-dwarf galaxy rich of gas and of dark matter, with the recently enhanced star formation responsible for its blue colour. As it was argued above, star formation is most probably the result of accretion of low metal abundant gas onto dwarf satellite and the NE-part of NGC4656.

A possible  source of accreting gas is the neighbor edge-on galaxy NGC4631, situated at a distance of about 60 kpc in the sky plane. Fingers of \HI emerging from NGC4631 including in the direction of NGC4656 are very prominent in the \HI distribution map around this galaxy   \citep[][]{Rand1994, Schechtman-Rooketal2012}. A total mass of \HI in the most prominent tidal spurs reaches $3\cdot 10^9 M_\odot$, which is almost half of the total mass of \HI inside of NGC4631.  Note that  NGC4631 is the moderately underabundant galaxy: according to \cite{Pilyuginetal2014}\footnote{ Pilyugin et al. (2014) estimated the oxygen abundance in NGC 4631 using C-calibration, which is consistent with the S-method used in our paper.}, the central value of oxygen abundance $12+\log\mathrm{(O/H)_S}$ is 8.39$\pm$ 0.06, and it  decreases down to 8.0 at the radius $R_{25}$. Hence, to explain the O/H abundance of accreted gas one can conclude that this gas was initially situated beyond the optical radius of the galaxy.  Taking into account the gradient of O/H, such metal poor gas should come out of the outermost parts of the galaxy $\sim 2.5 R_{25}$, which makes this origin of the accretion less realistic. Rather, if the gas had previously belonged to NGC4631, it was diluted by an intergalactic medium. However, similar role may play the accretion of  low-enriched gas from the extended gaseous discs of both NGC4656 and  its satellite. Gas, which envelops  the stellar bodies of these galaxies, is clearly seen in the \HI maps.  

Note that NGC4656UV strongly resembles  H~\textsc{i}-rich ultra-diffuse galaxies (UDGs) where star formation of low efficiency takes place, see \cite{Leismanetal2017}. A very similar object of this kind is the nearest galaxy of this type  UGC2172, which is twice as distant as NGC4656UV.  This starforming UDG was recently studied in details by \cite{Trujillo2017}.   Both UGC2172 and NGC4656UV have similar central brightness $\sim 24 ~mag/arcsec^2$,  close stellar masses $ \sim$ (1-4)$\cdot 10^7 M_\odot$ and gas content  $M_{HI}\sim (2-4)\cdot 10^8 M_\odot$. 

\begin{figure*}                                                              
 \centering     
\includegraphics[scale=0.3]{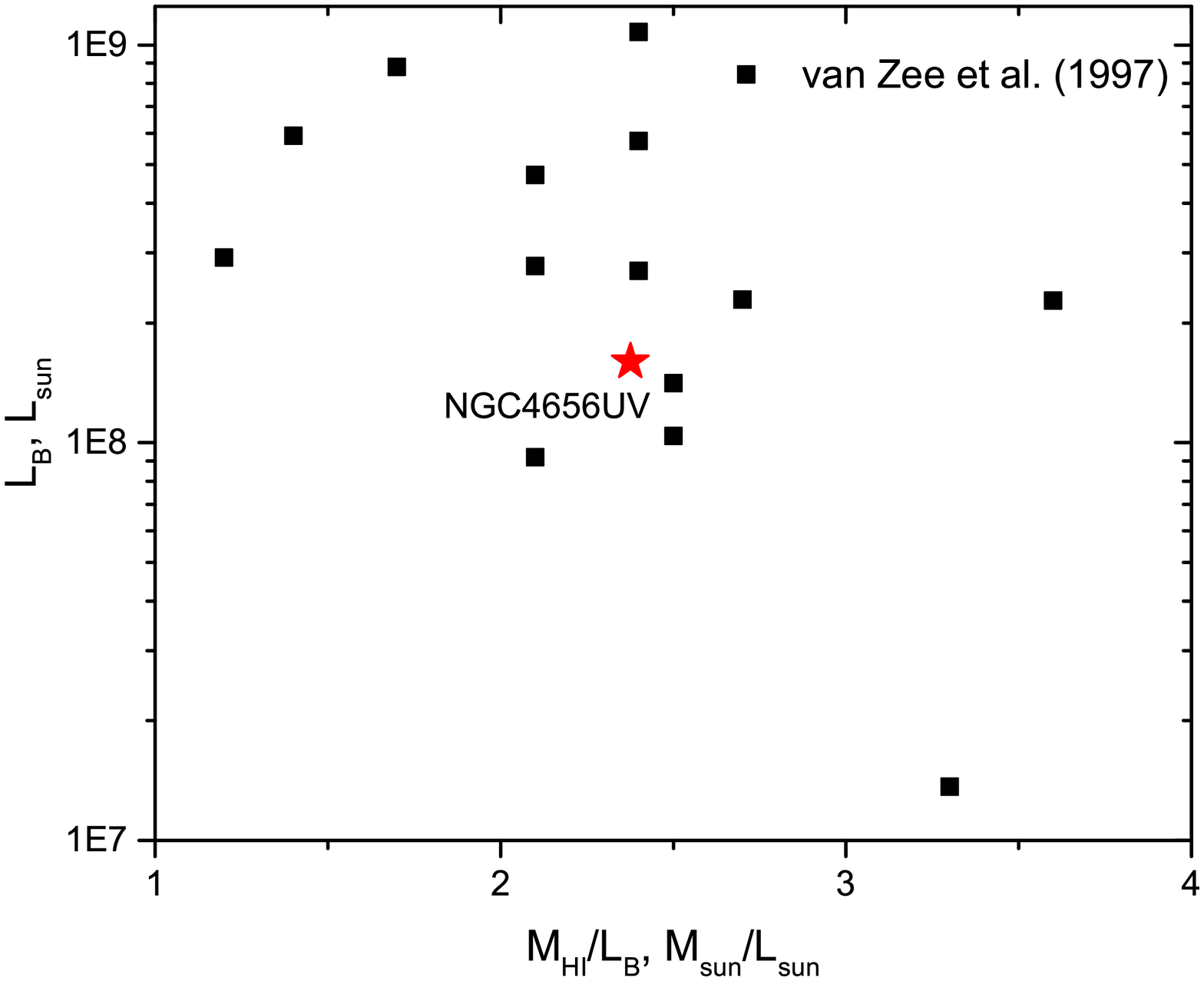}   
\includegraphics[scale=0.3]{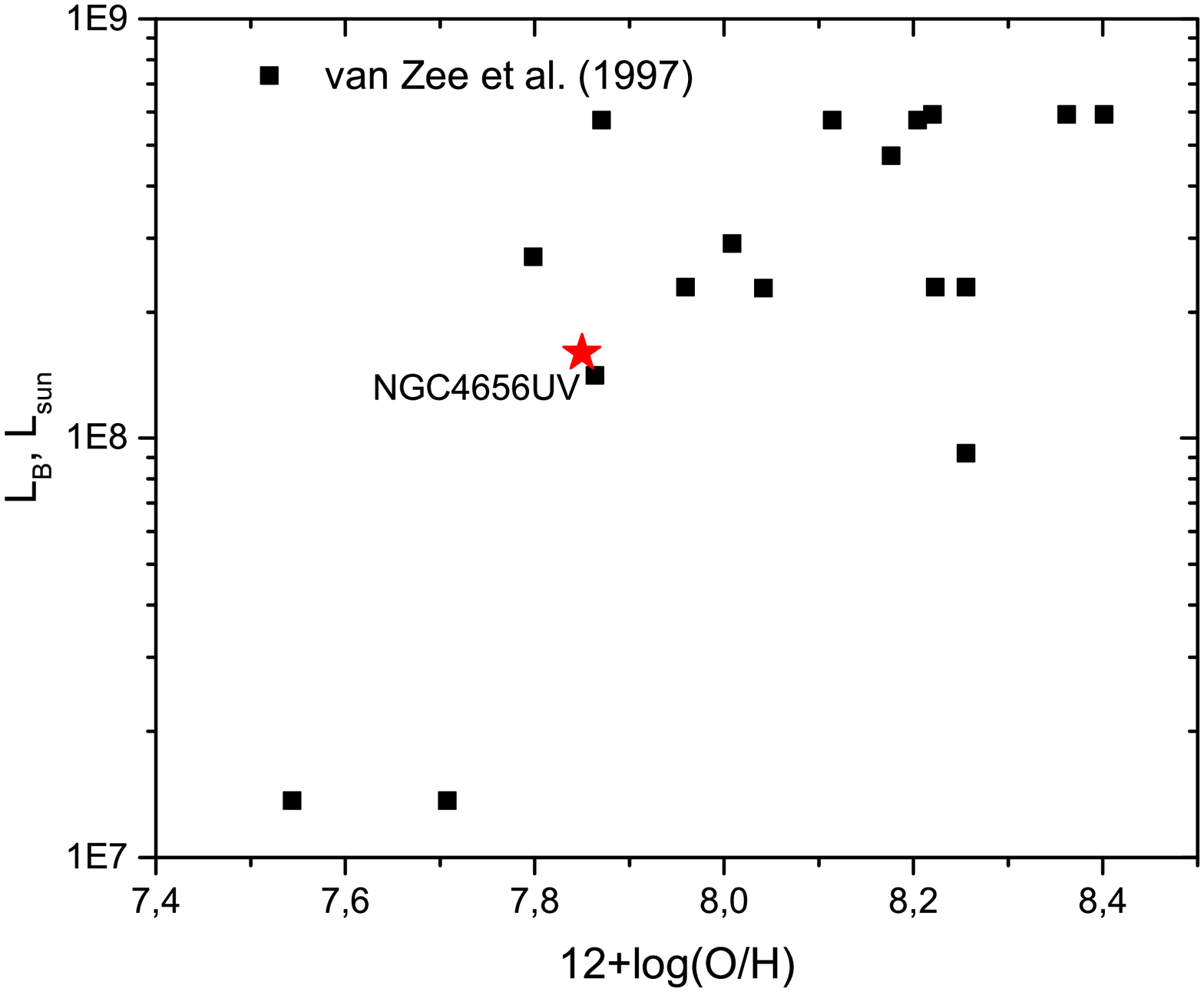}  
\caption{ A comparison of NGC4656UV with the other LSB dwarf galaxies from van Zee et al (1997). }
\label{van_Zee} 
\end{figure*}

In Fig. \ref{van_Zee} we  compare NGC4656UV by mass of \HI  and O/H ratio with gas-rich dwarf LSB galaxies of similar luminosities and central surface brightnesses taken from the sample of \cite{vanZee1997}. The position of NGC4656UV is shown by star, square symbols correspond to the data from \cite{vanZee1997}. We passed to B-band luminosity from  $L_r$ luminosity and colour indices using the transformation equations. We took the estimate of the \HI mass from \cite{Schechtman-Rooketal2012}.  The data of \cite{vanZee1997} were adjusted to the of Hubble constant h=75 $\textrm{km}~s^{-1}/Mpc$.  From the diagrams one can see that NGC4656UV   does not outstand from the other LSB dwarfs.

However, there are at least three peculiarities of this galaxy which make it different from the majority of the other LSB dwarfs, possessing similar luminosity and gas content. First, this galaxy is currently interacting with its more massive neighbour,  which explains the non-circular gas motions, as well as the recent star formation triggered by interaction and gas accretion. Second,   star formation in this galaxy reveals itself by non-resolved emission patches scattered allover the disc in the absence of the extended H~\textsc{ii}  complexes, usually observed in Irr-galaxies. It evidences a low density of gas with no large-scale instabilities,  when most of UV radiation leaks out from the gas layer. Similar character of star formation
is also often observed in the far outskirts of discs of gas-rich galaxies \citep[see f.e.][]{Werketal2010}. Third,  
NGC4656UV is unusually blue with respect to other dwarf galaxies, including UDGs: its color and spectral energy distribution corresponds to the luminosity-weighted age of low-abundant stellar population which does not exceed several hundred million years, although the presence of the old population cannot be excluded \citep{Schechtman-Rooketal2012}. Evidently, in the absence of recent star formation this small galaxy would hardly be noticeable at all in the optical bands.

\section{Conclusions}\label{conclusion} 
We studied the kinematics and oxygen abundance of galaxy NGC4656 and its neighbour UV-bright LSB dwarf galaxy NGC4656UV, based on the long-slit spectral observations.
 Our estimates of gas velocities of NGC4656UV parallel with photometrical and \HI data speak in favour of this system to be the dark matter-dominated gravitationally bound galaxy with the low surface brightness ($(\mu_{0})_r= 24.12 ~mag/arcsec^2$, non-corrected for disc inclination), rather than a  tidal dwarf candidate. The parameters of NGC4656UV are close to that observed for gas-rich dwarf LSB-galaxies and ultra-diffuse galaxies.   Note that in the absence of recent (or current) star formation it would be extremely hard to discover such  a low brightness object.

 Oxygen gas-phase abundances found for the brightest \HII-region of NGC4656UV as well as for the emission gas of the main galaxy NGC4656 at the side facing UV dwarf,  are low: 12+log O/H = 7.8- 8.1 (\textit{izi}-method). Both the low abundance and non-circular gas motions in NGC4656 parallel with the observed young stellar population of its dwarf satellite NGC4656UV are considered as the evidences of the current accretion of metal-poor gas on the discs of both galaxies as the result of tidal interaction.

\section*{Acknowledgements} 
The authors thank the anonymous referee for valuable comments that helped to improve the paper. 
The authors acknowledge the support of RFBR grants  14-22-03006, 15-52-15050  (observations) and the Russian Science Foundation (RSCF) grant 14-22-00041 (data processing and analysis). This
research has made use of the Lyon Extragalactic Database (LEDA,
http://leda.univ-lyon1.fr). In this study, we used the SDSS
DR13 data. Funding for the SDSS and SDSS-II has been pro-
vided by the Alfred P. Sloan Foundation, the Participating Insti-
tutions, the National Science Foundation, the U.S. Department of
Energy, the National Aeronautics and Space Administration, the
Japanese Monbukagakusho, the Max Planck Society, and the Higher
Education Funding Council for England. The SDSS Web site is
http://www.sdss.org/. The Russian 6-m telescope is exploited under the financial support by the Russian Federation Ministry of Education and Science (agreement No14.619.21.0004, project ID RFMEFI61914X0004).

\bibliographystyle{mnras}
\bibliography{n4656}

\begin{thebibliography}{}
\makeatletter
\relax
\def\mn@urlcharsother{\let\do\@makeother \do\$\do\&\do\#\do\^\do\_\do\%\do\~}
\def\mn@doi{\begingroup\mn@urlcharsother \@ifnextchar [ {\mn@doi@}
  {\mn@doi@[]}}
\def\mn@doi@[#1]#2{\def\@tempa{#1}\ifx\@tempa\@empty \href
  {http://dx.doi.org/#2} {doi:#2}\else \href {http://dx.doi.org/#2} {#1}\fi
  \endgroup}
\def\mn@eprint#1#2{\mn@eprint@#1:#2::\@nil}
\def\mn@eprint@arXiv#1{\href {http://arxiv.org/abs/#1} {{\tt arXiv:#1}}}
\def\mn@eprint@dblp#1{\href {http://dblp.uni-trier.de/rec/bibtex/#1.xml}
  {dblp:#1}}
\def\mn@eprint@#1:#2:#3:#4\@nil{\def\@tempa {#1}\def\@tempb {#2}\def\@tempc
  {#3}\ifx \@tempc \@empty \let \@tempc \@tempb \let \@tempb \@tempa \fi \ifx
  \@tempb \@empty \def\@tempb {arXiv}\fi \@ifundefined
  {mn@eprint@\@tempb}{\@tempb:\@tempc}{\expandafter \expandafter \csname
  mn@eprint@\@tempb\endcsname \expandafter{\@tempc}}}

\bibitem[\protect\citeauthoryear{{Afanasiev} \& {Moiseev}}{{Afanasiev} \&
  {Moiseev}}{2005}]{AfanasievMoiseev2005}
{Afanasiev} V.~L.,  {Moiseev} A.~V.,  2005, \mn@doi [Astronomy Letters]
  {10.1134/1.1883351}, \href
  {http://adsabs.harvard.edu/abs/2005AstL...31..194A} {31, 194}

\bibitem[\protect\citeauthoryear{{Afanasiev} \& {Moiseev}}{{Afanasiev} \&
  {Moiseev}}{2011}]{AfanasievMoiseev2011}
{Afanasiev} V.~L.,  {Moiseev} A.~V.,  2011, Baltic Astronomy, \href
  {http://adsabs.harvard.edu/abs/2011BaltA..20..363A} {20, 363}

\bibitem[\protect\citeauthoryear{{Baldwin}, {Phillips}  \&
  {Terlevich}}{{Baldwin} et~al.}{1981}]{BPT}
{Baldwin} J.~A.,  {Phillips} M.~M.,   {Terlevich} R.,  1981, \mn@doi [\pasp]
  {10.1086/130766}, \href {http://adsabs.harvard.edu/abs/1981PASP...93....5B}
  {93, 5}

\bibitem[\protect\citeauthoryear{{Blanc}, {Kewley}, {Vogt}  \&
  {Dopita}}{{Blanc} et~al.}{2015}]{izi}
{Blanc} G.~A.,  {Kewley} L.,  {Vogt} F.~P.~A.,   {Dopita} M.~A.,  2015, \mn@doi
  [\apj] {10.1088/0004-637X/798/2/99}, \href
  {http://adsabs.harvard.edu/abs/2015ApJ...798...99B} {798, 99}

\bibitem[\protect\citeauthoryear{{Chilingarian}, {Prugniel}, {Sil'Chenko}  \&
  {Afanasiev}}{{Chilingarian} et~al.}{2007}]{Chilingarian2007}
{Chilingarian} I.~V.,  {Prugniel} P.,  {Sil'Chenko} O.~K.,   {Afanasiev} V.~L.,
   2007, \mn@doi [\mnras] {10.1111/j.1365-2966.2007.11549.x}, \href
  {http://adsabs.harvard.edu/abs/2007MNRAS.376.1033C} {376, 1033}

\bibitem[\protect\citeauthoryear{{Combes}}{{Combes}}{1978}]{Combes1978}
{Combes} F.,  1978, \aap, \href
  {http://adsabs.harvard.edu/abs/1978A%26A....65...47C} {65, 47}

\bibitem[\protect\citeauthoryear{{Donahue}, {Aldering}  \& {Stocke}}{{Donahue}
  et~al.}{1995}]{Donahue1995}
{Donahue} M.,  {Aldering} G.,   {Stocke} J.~T.,  1995, \mn@doi [\apjl]
  {10.1086/316771}, \href {http://adsabs.harvard.edu/abs/1995ApJ...450L..45D}
  {450, L45}

\bibitem[\protect\citeauthoryear{{Into} \& {Portinari}}{{Into} \&
  {Portinari}}{2013}]{IntoPortinari2013}
{Into} T.,  {Portinari} L.,  2013, \mn@doi [\mnras] {10.1093/mnras/stt071},
  \href {http://adsabs.harvard.edu/abs/2013MNRAS.430.2715I} {430, 2715}

\bibitem[\protect\citeauthoryear{{Karachentsev}, {Makarov}  \&
  {Kaisina}}{{Karachentsev} et~al.}{2013}]{Karachentsev2013}
{Karachentsev} I.~D.,  {Makarov} D.~I.,   {Kaisina} E.~I.,  2013, \mn@doi [\aj]
  {10.1088/0004-6256/145/4/101}, \href
  {http://adsabs.harvard.edu/abs/2013AJ....145..101K} {145, 101}

\bibitem[\protect\citeauthoryear{{Karachentsev}, {Bautzmann}, {Neyer}, {Polzl},
  {Riepe}, {Zilch}  \& {Mattern}}{{Karachentsev}
  et~al.}{2014}]{Karachentsevetal2014}
{Karachentsev} I.~D.,  {Bautzmann} D.,  {Neyer} F.,  {Polzl} R.,  {Riepe} P.,
  {Zilch} T.,   {Mattern} B.,  2014, preprint, \href
  {http://adsabs.harvard.edu/abs/2014arXiv1401.2719K} {} (\mn@eprint {arXiv}
  {1401.2719})

\bibitem[\protect\citeauthoryear{{Kasparova}, {Saburova}, {Katkov},
  {Chilingarian}  \& {Bizyaev}}{{Kasparova} et~al.}{2014}]{Kasparova2014}
{Kasparova} A.~V.,  {Saburova} A.~S.,  {Katkov} I.~Y.,  {Chilingarian} I.~V.,
  {Bizyaev} D.~V.,  2014, \mn@doi [\mnras] {10.1093/mnras/stt1982}, \href
  {http://adsabs.harvard.edu/abs/2014MNRAS.437.3072K} {437, 3072}

\bibitem[\protect\citeauthoryear{{Kauffmann} et~al.,}{{Kauffmann}
  et~al.}{2003}]{kauffmann03}
{Kauffmann} G.,  et~al., 2003, \mn@doi [\mnras]
  {10.1111/j.1365-2966.2003.07154.x}, \href
  {http://adsabs.harvard.edu/abs/2003MNRAS.346.1055K} {346, 1055}

\bibitem[\protect\citeauthoryear{{Kewley} \& {Ellison}}{{Kewley} \&
  {Ellison}}{2008}]{kewley08}
{Kewley} L.~J.,  {Ellison} S.~L.,  2008, \mn@doi [\apj] {10.1086/587500}, \href
  {http://adsabs.harvard.edu/abs/2008ApJ...681.1183K} {681, 1183}

\bibitem[\protect\citeauthoryear{{Kewley}, {Dopita}, {Sutherland}, {Heisler}
  \& {Trevena}}{{Kewley} et~al.}{2001}]{kewley2001}
{Kewley} L.~J.,  {Dopita} M.~A.,  {Sutherland} R.~S.,  {Heisler} C.~A.,
  {Trevena} J.,  2001, \mn@doi [\apj] {10.1086/321545}, \href
  {http://adsabs.harvard.edu/abs/2001ApJ...556..121K} {556, 121}

\bibitem[\protect\citeauthoryear{{Le Borgne}, {Rocca-Volmerange}, {Prugniel},
  {Lan{\c c}on}, {Fioc}  \& {Soubiran}}{{Le Borgne}
  et~al.}{2004}]{LeBorgne2004}
{Le Borgne} D.,  {Rocca-Volmerange} B.,  {Prugniel} P.,  {Lan{\c c}on} A.,
  {Fioc} M.,   {Soubiran} C.,  2004, \mn@doi [\aap]
  {10.1051/0004-6361:200400044}, \href
  {http://adsabs.harvard.edu/abs/2004A%26A...425..881L} {425, 881}

\bibitem[\protect\citeauthoryear{{Leisman} et~al.,}{{Leisman}
  et~al.}{2017}]{Leismanetal2017}
{Leisman} L.,  et~al., 2017, preprint, \href
  {http://adsabs.harvard.edu/abs/2017arXiv170305293L} {} (\mn@eprint {arXiv}
  {1703.05293})

\bibitem[\protect\citeauthoryear{{L{\'o}pez-S{\'a}nchez}, {Dopita}, {Kewley},
  {Zahid}, {Nicholls}  \& {Scharw{\"a}chter}}{{L{\'o}pez-S{\'a}nchez}
  et~al.}{2012}]{lopez-sanchez12}
{L{\'o}pez-S{\'a}nchez} {\'A}.~R.,  {Dopita} M.~A.,  {Kewley} L.~J.,  {Zahid}
  H.~J.,  {Nicholls} D.~C.,   {Scharw{\"a}chter} J.,  2012, \mn@doi [\mnras]
  {10.1111/j.1365-2966.2012.21145.x}, \href
  {http://adsabs.harvard.edu/abs/2012MNRAS.426.2630L} {426, 2630}

\bibitem[\protect\citeauthoryear{{Makarov}, {Prugniel}, {Terekhova}, {Courtois}
   \& {Vauglin}}{{Makarov} et~al.}{2014}]{Makarovetal2014}
{Makarov} D.,  {Prugniel} P.,  {Terekhova} N.,  {Courtois} H.,   {Vauglin} I.,
  2014, \mn@doi [\aap] {10.1051/0004-6361/201423496}, \href
  {http://adsabs.harvard.edu/abs/2014A%26A...570A..13M} {570, A13}

\bibitem[\protect\citeauthoryear{{Mart{\'{\i}}nez-Delgado}, {D'Onghia},
  {Chonis}, {Beaton}, {Teuwen}, {GaBany}, {Grebel}  \&
  {Morales}}{{Mart{\'{\i}}nez-Delgado} et~al.}{2015}]{Martinez-Delgadoetal2015}
{Mart{\'{\i}}nez-Delgado} D.,  {D'Onghia} E.,  {Chonis} T.~S.,  {Beaton} R.~L.,
   {Teuwen} K.,  {GaBany} R.~J.,  {Grebel} E.~K.,   {Morales} G.,  2015,
  \mn@doi [\aj] {10.1088/0004-6256/150/4/116}, \href
  {http://adsabs.harvard.edu/abs/2015AJ....150..116M} {150, 116}

\bibitem[\protect\citeauthoryear{{Pilyugin} \& {Grebel}}{{Pilyugin} \&
  {Grebel}}{2016}]{pilyugin16}
{Pilyugin} L.~S.,  {Grebel} E.~K.,  2016, \mn@doi [\mnras]
  {10.1093/mnras/stw238}, \href
  {http://adsabs.harvard.edu/abs/2016MNRAS.457.3678P} {457, 3678}

\bibitem[\protect\citeauthoryear{{Pilyugin}, {Grebel}  \& {Kniazev}}{{Pilyugin}
  et~al.}{2014}]{Pilyuginetal2014}
{Pilyugin} L.~S.,  {Grebel} E.~K.,   {Kniazev} A.~Y.,  2014, \mn@doi [\aj]
  {10.1088/0004-6256/147/6/131}, \href
  {http://adsabs.harvard.edu/abs/2014AJ....147..131P} {147, 131}

\bibitem[\protect\citeauthoryear{{Rand}}{{Rand}}{1994}]{Rand1994}
{Rand} R.~J.,  1994, \aap, \href
  {http://adsabs.harvard.edu/abs/1994A%26A...285..833R} {285, 833}

\bibitem[\protect\citeauthoryear{{Roediger} \& {Courteau}}{{Roediger} \&
  {Courteau}}{2015}]{RoedigerCourteau2015}
{Roediger} J.~C.,  {Courteau} S.,  2015, \mn@doi [\mnras]
  {10.1093/mnras/stv1499}, \href
  {http://adsabs.harvard.edu/abs/2015MNRAS.452.3209R} {452, 3209}

\bibitem[\protect\citeauthoryear{{Schechtman-Rook} \& {Hess}}{{Schechtman-Rook}
  \& {Hess}}{2012}]{Schechtman-Rooketal2012}
{Schechtman-Rook} A.,  {Hess} K.~M.,  2012, \mn@doi [\apj]
  {10.1088/0004-637X/750/2/171}, \href
  {http://adsabs.harvard.edu/abs/2012ApJ...750..171S} {750, 171}

\bibitem[\protect\citeauthoryear{{Seth}, {Dalcanton}  \& {de Jong}}{{Seth}
  et~al.}{2005}]{Sethetal2005}
{Seth} A.~C.,  {Dalcanton} J.~J.,   {de Jong} R.~S.,  2005, \mn@doi [\aj]
  {10.1086/427859}, \href {http://adsabs.harvard.edu/abs/2005AJ....129.1331S}
  {129, 1331}

\bibitem[\protect\citeauthoryear{{Trujillo}, {Roman}, {Filho}  \& {S{\'a}nchez
  Almeida}}{{Trujillo} et~al.}{2017}]{Trujillo2017}
{Trujillo} I.,  {Roman} J.,  {Filho} M.,   {S{\'a}nchez Almeida} J.,  2017,
  \mn@doi [\apj] {10.3847/1538-4357/aa5cbb}, \href
  {http://adsabs.harvard.edu/abs/2017ApJ...836..191T} {836, 191}

\bibitem[\protect\citeauthoryear{{Werk} et~al.,}{{Werk}
  et~al.}{2010}]{Werketal2010}
{Werk} J.~K.,  et~al., 2010, \mn@doi [\aj] {10.1088/0004-6256/139/1/279}, \href
  {http://adsabs.harvard.edu/abs/2010AJ....139..279W} {139, 279}

\bibitem[\protect\citeauthoryear{{Zasov}, {Saburova}, {Katkov}, {Egorov}  \&
  {Afanasiev}}{{Zasov} et~al.}{2015}]{zasovetal2015}
{Zasov} A.,  {Saburova} A.,  {Katkov} I.,  {Egorov} O.,   {Afanasiev} V.,
  2015, \mn@doi [\mnras] {10.1093/mnras/stv454}, \href
  {http://adsabs.harvard.edu/abs/2015MNRAS.449.1605Z} {449, 1605}

\bibitem[\protect\citeauthoryear{{Zasov}, {Saburova}, {Egorov}  \&
  {Afanasiev}}{{Zasov} et~al.}{2016}]{zasovetal2016}
{Zasov} A.~V.,  {Saburova} A.~S.,  {Egorov} O.~V.,   {Afanasiev} V.~L.,  2016,
  \mn@doi [\mnras] {10.1093/mnras/stw1905}, \href
  {http://adsabs.harvard.edu/abs/2016MNRAS.462.3419Z} {462, 3419}

\bibitem[\protect\citeauthoryear{{van Zee}, {Haynes}  \& {Salzer}}{{van Zee}
  et~al.}{1997}]{vanZee1997}
{van Zee} L.,  {Haynes} M.~P.,   {Salzer} J.~J.,  1997, \mn@doi [\aj]
  {10.1086/118662}, \href {http://adsabs.harvard.edu/abs/1997AJ....114.2497V}
  {114, 2497}

\makeatother
\end{thebibliography}

\label{lastpage}

\end{document}